\documentclass[aoas,nameyear,seceqn,dvips]{arximspdf}
\usepackage{multirow,dcolumn,stfloats}
\usepackage{graphicx}

\doi{10.1214/10-AOAS344}
\volume{4}
\issue{4}
\pubyear{2010}
\firstpage{2073}
\lastpage{2098}

\makeatletter

\fnbelowfloat
\DeclareMathAlphabet{\boldmathcaligr}{OMS}{cmsy}{b}{n}
\newcolumntype{d}[1]{D{.}{.}{#1}}
\newtheorem{ass}{Assumption}[section]
\newtheorem{lem}{Lemma}[section]
\newtheorem{Thm}{Theorem}[section]

\newcommand{\Num}{\operatorname{Num}}
\newcommand{\Prr}{\operatorname{Pr}}
\newcommand{\dd}{\mathrm{d}}
\newcommand{\vecc}{\mathrm{vec}}
\newcommand{\tptwo}{\mathrm{TP}_2}
\newcommand{\mtptwo}{\mathrm{MTP}_2}
\newcommand{\cov}{\operatorname{cov}}

\newcommand{\bbB}{\mathbb{B}}
\newcommand{\bbI}{\mathbb{I}}
\newcommand{\Int}{\mathbb{I}}
\newcommand{\bbN}{\mathbb{N}}
\newcommand{\Nat}{\mathbb{N}}
\newcommand{\Real}{\mathbb{R}}
\newcommand{\bbS}{\mathbb{S}}
\newcommand{\bbT}{\mathbb{T}}

\newcommand{\lows}{\underline{s}}
\newcommand{\upps}{\bar{s}}
\newcommand{\bzero}{\mathbf{0}}

\newcommand{\bE}{\mathbf{E}}
\newcommand{\bH}{\mathbf{H}}
\newcommand{\bP}{\mathbf{P}}

\newcommand{\bPi}{\bolds{\Pi}}

\newcommand{\bmi}{\mathbf{i}}
\newcommand{\bmu}{\mathbf{u}}

\newcommand{\bcI}{\boldmathcaligr{I}}

\newcommand{\bdelta}{\bolds{\delta}}
\newcommand{\biota}{\bolds{\iota}}
\newcommand{\bsigma}{\bolds{\sigma}}

\newcommand{\bmS}{\mathbf{S}}
\newcommand{\bmV}{\mathbf{V}}
\newcommand{\bmY}{\mathbf{Y}}

\newcommand{\C}{\mathbb{C}}

\newcommand{\cA}{\mathcal{A}}
\newcommand{\cB}{\mathcal{B}}
\newcommand{\cC}{\mathcal{C}}
\newcommand{\cD}{\mathcal{D}}
\newcommand{\cE}{\mathcal{E}}
\newcommand{\cL}{\mathcal{L}}
\newcommand{\cM}{\mathcal{M}}
\newcommand{\cP}{\mathcal{P}}
\newcommand{\cS}{\mathcal{S}}

\makeatother

\begin{document}
\begin{frontmatter}

\title{Testing affiliation in private-values models of first-price
auctions using grid distributions}
\runtitle{Testing affiliation in models of auctions}

\begin{aug}
\author[A]{\fnms{Luciano I.} \snm{de Castro}\ead[label=e1]{deCastro.Luciano@gmail.com}}
\and
\author[B]{\fnms{Harry J.} \snm{Paarsch}\corref{}\ead[label=e2]{HPaarsch@UniMelb.edu.au}}

\runauthor{L. I. de Castro and H. J. Paarsch}
\affiliation{Northwestern University and University of Melbourne}
\address[A]{MEDS, Kellogg School of Management\\
Northwestern University\\
USA\\
\printead{e1}} 
\address[B]{Department of Economics\\
University of Melbourne\\
Australia\\
\printead{e2}}
\end{aug}

\received{\smonth{5} \syear{2009}}
\revised{\smonth{3} \syear{2010}}

%
\begin{abstract}
Within the private-values paradigm, we construct a tractable empirical model
of equilibrium behavior at first-price auctions when bidders'
valuations are
potentially dependent, but not necessarily affiliated. We develop a
test of
affiliation and apply our framework to data from low-price, sealed-bid auctions
held by the Department of Transportation in the State of Michigan to procure
road-resurfacing services: we do not reject the hypothesis of
affiliation in
cost signals.
\end{abstract}


\end{frontmatter}

\section{Motivation and introduction}\label{s1}

During the past half century, economists have made considerable
progress in
understanding the theoretical structure of equilibrium strategic behavior
under market mechanisms, such as auctions, when the number of potential
participants is relatively small; see \citet{krishna2010} for a comprehensive
presentation and evaluation of progress.

One analytic device, commonly used to describe bidder motivation at
single-object auctions, is a continuous random variable which represents
individual-specific heterogeneity in valuations. The conceptual experiment
involves each \mbox{potential} bidder's receiving a draw from a distribution of
valuations. Conditional on his draw, a bidder is then assumed to act
purposefully, maximizing either the expected profit or the expected
utility of
profit from winning the auction. Another frequently-made assumption is that
the valuation draws of bidders are independent and that the bidders are
\textit{ex ante} symmetric---their draws being from the same distribution of valuations.
This framework is often referred to as the \textit{symmetric independent
private-values paradigm} (symmetric IPVP). Under these assumptions, a
researcher can then focus on a representative agent's decision rule when
describing equilibrium behavior.

At many real-world auctions, the latent valuations of potential bidders are
probably dependent in some way. In auction theory, it has been assumed that
dependence satisfies \textit{affiliation}, a term coined by \citet{milgwebe1982}.
Affiliation is a condition concerning the joint distribution of signals.
Often, affiliation is described using the intuition presented by Milgrom
and Weber: ``roughly, this [affiliation] means that a high value of one
bidder's estimate makes high values of the others' estimates more likely.''
Thus described, affiliation seems like a relatively innocuous
condition. In
the case of continuous random variables, following the path blazed by
\citet{karlin1968}, some refer to affiliation as \textit{multivariate total
positivity of order two}, or $\mtptwo$ for short. Essentially, under
affiliation, with continuous random variables, the off-diagonal
elements of
the Hessian of the logarithm of the joint probability density function of
signals are all nonnegative, that is, the joint probability density
function is
log-supermodular. Under joint normality of signals, affiliation requires
that \textit{all} the pair-wise covariances be weakly positive.

How is affiliation related to other forms of dependence? Consider two
continuous random variables $V_1$ and $V_2$, having joint probability density
function $f_{V_1,V_2}(v_1,v_2)$ as well as conditional probability density
functions $f_{V_2|V_1}(v_2|v_1)$ and $f_{V_1|V_2}(v_1|v_2)$ and conditional
cumulative distribution functions $F_{V_2|V_1}(v_2|v_1)$ and
$F_{V_1|V_2}(v_1|v_2)$. Introduce $g(\cdot)$ and $h(\cdot)$,
functions that
are nondecreasing in their arguments. \citet{decastro2007} has noted that
affiliation implies (a)~$[F_{V_2|V_1}(v_2| v_1)/f_{V_2|V_1}(v_2|v_1)]$ is
decreasing in $v_1$ (and $v_2$ in the other case), often referred to as a
decreasing \textit{inverse hazard rate}, which implies (b)
$\Prr(V_2\le v_2|V_1=v_1)$ is nonincreasing in $v_1$ (and $v_2$ in the other
case), also referred to as \textit{positive regression dependence},
which implies
(c) $\Prr(V_2\le v_2|V_1\le v_1)$ is nonincreasing in $v_1$ (and $v_2$ in
the other case), also referred to as \textit{left-tail decreasing} in $v_1$
($v_2$), which implies (d) $\cov[g(V_1,V_2),h(V_1,V_2)]$ is positive, which
implies that (e) $\cov[g(V_1),h(V_2)]$ is positive, which implies (f)
$\cov(V_1,V_2)$ is positive. The important point to note is that affiliation
is a much stronger form of dependence than positive covariance. In addition,
\citet{decastro2007} has demonstrated that, within the set of all signal
distributions, the set satisfying affiliation is small, both in the topological
sense and in the measure-theoretic sense.

Affiliation delivers several predictions and results: first, under affiliation,
the existence and uniqueness of a monotone pure-strategy equilibrium
(MPSE) is
guaranteed. Also, four commonly-studied auction formats---the
oral,\break
ascending-price (often referred to by economists as the \textit{English}) and
the second-price, sealed-bid (often referred to by economists as the
\textit{Vickrey}) as well as two first-price ones---can be ranked in
terms of the
revenues they can be expected to generate. Specifically, the expected revenues
at English auctions are weakly greater than those at Vickrey auctions
which are
greater than those at first-price auctions---either the sealed-bid or
the oral,
descending-price (often referred to by economists as the \textit{Dutch}) formats.
Note, however, that when bidders are asymmetric, their valuation draws being
from different marginal distributions, these rankings no longer apply. In
fact, in general, very little can be said about the revenue-generating
properties of the various auction formats and pricing rules under asymmetries.

Empirically investigating equilibrium behavior at auctions when latent
valuations are affiliated has challenged researchers for some time.
\citet{laffvuon1996} showed that identification has been impossible to
establish in many models when affiliation is present. In fact, Laffont and
Vuong demonstrated that any model within the affiliated-values paradigm (AVP)
is observationally equivalent to a model within the affiliated private-values
paradigm (APVP). For this reason, virtually all empirical workers who have
considered some form of dependence have worked within the APVP.

Only a few researchers have dealt explicitly with models within the APVP.
In particular, \citet{lietal2000} have demonstrated nonparametric
identification within the conditional IPVP, a special case of the APVP,
while \citet{lietal2002} have demonstrated nonparametric identification
within the APVP. One of the problems that Li et al. faced when implementing
their approach is that nonparametric kernel-smoothed estimators are
often slow
to converge. In addition, Li et al. do not impose affiliation in their
estimation strategy, so the first-order condition used in their two-step
estimation strategy need not constitute an equilibrium. \citet{hubbetal2009}
have sought to address some of these technical problems using semiparametric
methods which sacrifice the full generality of the nonparametric approach
\textit{in lieu} of additional structure.

To date, except for \citet{brenpaar2007}, no one has attempted to examine,
empirically, models in which the private values are potentially
dependent, but
not necessarily affiliated. Incidentally, using data from sequential English
auctions of two different objects, Brendstrup and Paarsch found weak evidence
against affiliation in the valuation draws of two objects for the same bidder.

\citet{decastro2007} has noted that, within the private-values paradigm,
affiliation is unnecessary to guarantee the existence and uniqueness of a
MPSE. In fact, he has demonstrated existence and uniqueness of a MPSE under
a weaker form of dependence, one where the inverse hazard rate is decreasing
in the conditioned argument.

Because affiliation is unnecessary to guarantee existence and
uniqueness of
bidding strategies in models of first-price auctions with private values,
expected revenue predictions based on empirical models in which affiliation
is imposed are potentially biased. Knowing whether valuations are affiliated
is central to ranking auction formats in terms of the expected revenues
generated. In the absence of affiliation, the expected-revenue rankings
delivered by the \textit{linkage principle} of \citet{milgwebe1982} need
not hold: the expected-revenue rankings across auctions formats remain an
empirical question. Thus, investigating the empirical validity of affiliation
appears both an important and a useful exercise.

In next section of this paper we present a brief description of affiliation
and its soldier---total positivity of order two ($\tptwo$). Subsequently,
following the theoretical work of de Castro (\citeyear{decastro2007}, \citeyear{decastro2008}), who
introduced the notion of the grid distribution, in Section \ref{s3} we
construct a
tractable empirical model of equilibrium behavior at first-price
auctions when
the private valuations of bidders are potentially dependent, but not necessarily
affiliated.\footnote{The grid distributions discussed and used in this paper can also
be modeled as contingency tables, which have been used extensively
in applications in the social sciences; see \citet{dougetal1990}
for the connections between contingency tables and positive
dependence properties, including affiliation ($\tptwo$), which
is the focus of this paper.}
In Section \ref{s4} we develop a test of affiliation, which is based on
grid distributions, rather than kernel-smoothing methods, thus avoiding the
drawback encountered by Li, Perrigne and Vuong (\citeyear{lietal2000}, \citeyear{lietal2002}), while in
Section \ref{s5}
we apply our methods in an empirical investigation of low-price, sealed-bid,
procurement-contract auctions held by the Department of Transportation in
the State of Michigan, and do not reject the null hypothesis of affiliation.

This information is potentially useful to a policy maker. The apparent high
degree of estimated affiliation also explains why low levels of observed
competition are often sufficient to maintain relatively low profit margins:
strong affiliation is akin to fierce competition. Under strong
affiliation, a
potential winner knows that his nearest competitor probably has a valuation
(cost) close to his, and this disciplines his bidding behavior: he becomes
more aggressive than under independence. We summarize and conclude in Section
\ref{s6}, the final section of the paper. The results of a small-scale Monte
Carlo to
investigate the numerical as well as small-sample properties of our proposed
test are reported in the supplemental document---\citet{decapaar2010}.

\section{Affiliation and $\mathbf{TP_2}$}\label{s2}

\begin{figure}

\includegraphics{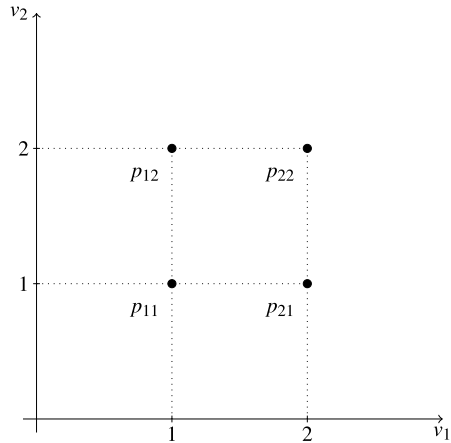}

  \caption{Affiliation with two bidders and two values for signals.}\label{fig:21}
\end{figure}

\begin{figure}

\includegraphics{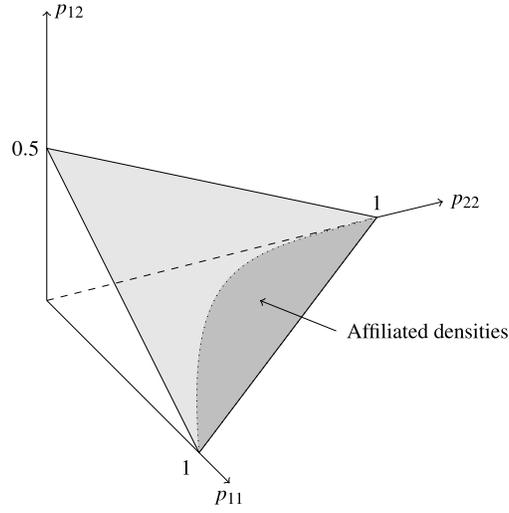}

\caption{Probability set: affiliation and alternative.}\label{fig:22}
\end{figure}

As mentioned above, affiliation is often described using an example with
two random variables that can take on either a low or a high value. The
two random variables are affiliated if high (low) values of each are more
likely to occur than high and low or low and high values. A commonly-used
graph of the four possible outcomes in a two-bidder auction game with two
values is depicted in Figure \ref{fig:21}. The $(1,1)$ and $(2,2)$ points
are more likely than the $(2,1)$ or $(1,2)$ points. Letting $p_{ij}$ denote
the probability of $(i,j)$, affiliation in this example then reduces to
$\tptwo$---viz.,
\[
p_{11}p_{22}\ge p_{12}p_{21}.
\]
Put another way, $\tptwo$ means that the determinant of the matrix
\[
\bP=
\pmatrix{
p_{11}&p_{12}\cr
p_{21}&p_{22}}
\]
must be weakly positive. Independence in valuations obviously satisfies the
lower bound on this determinental inequality. Note, too, that affiliation
restricts distributions to a part of the simplex depicted in Figure
\ref{fig:22}. In that figure, it is the region of the simplex that appears
below a semi-circle rising from the line where $p_{11}+p_{22}$ equals one.
In order to draw this figure, we needed to impose symmetry, so $p_{12}$ and
$p_{21}$ are equal; thus, the intercept for $p_{12}$ is one half. Conditions
that are weaker than affiliation, but that also guarantee existence and
uniqueness of equilibrium, are depicted in Figure \ref{fig:22}, too.
In fact,
in this simple example, the entire simplex satisfies these weaker conditions.
In richer examples, however, it is a subset of the simplex, but one that
contains the set of affiliated distributions. Thus, the assumption of
affiliation could be important in determining the revenues a seller can expect
from a particular auction format.

\citet{slavfien2009} have discussed geometric representations of
$2\times2$
distributions, like some of those considered here. Their
representations are
based on tetrahedrons, while ours reduce to triangles because of symmetry.

\begin{figure}

\includegraphics{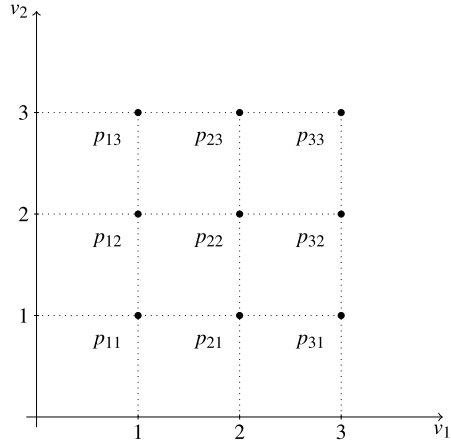}

\caption{Affiliation with two bidders and three values for signals.}\label{fig:23}
\end{figure}

Another important point to note is that affiliation is a global
restriction. To
see the importance of this fact, introduce the valuation $3$ for each bidder;
five additional points then appear, as is depicted in Figure \ref{fig:23}.
Affiliation requires that the probabilities at all collections of four points
satisfy $\tptwo$; that is, the following additional six inequalities
must hold:
\begin{eqnarray*}
p_{12}p_{23}&\ge &p_{13}p_{22},\qquad p_{22}p_{33}\ge p_{23}p_{32},\qquad p_{21}p_{32}\ge p_{22}p_{31},\\
p_{11}p_{33}&\ge &p_{13}p_{31},\qquad p_{12}p_{33}\ge p_{13}p_{32}\quad\mbox{and}\quad p_{21}p_{33}\ge p_{23}p_{31}.
\end{eqnarray*}
Of course, symmetry would imply that $p_{ij}$ equal $p_{ji}$ for all
$i$ and
$j$, so the joint mass function for two bidders and three valuations under
symmetric affiliation can be written as the following matrix:
\[
\bP=
\pmatrix{
p_{11}&p_{12}&p_{13}\cr
p_{21}&p_{22}&p_{23}\cr
p_{31}&p_{32}&p_{33}}=
\pmatrix{
a&d&e\cr
d&b&f\cr
e&f&c},
\]
where the determinants of all $(2\times2)$ submatrices must be positive.
Note, too, that all the points must also live on the simplex, so
\[
0\le a,b,c,d,e,f<1\quad\mbox{and}\quad a+b+c+2d+2e+2f = 1.
\]

How many inequalities are relevant? Let us represent the above matrix
in the
following tableau:
\[
\begin{tabular}{@{}clccccccc@{}}
\hline\\[-10pt]
\multirow{4}{0.5pt}[4.8pt]{\rule{0.5pt}{58pt}}&&\multirow{4}{0.5pt}[4.8pt]{\rule{0.5pt}{58pt}}
&1$\vphantom{0^{0^0}}$&\multirow{4}{0.5pt}[4.8pt]{\rule{0.5pt}{58pt}}
&2&\multirow{4}{0.5pt}[4.8pt]{\rule{0.5pt}{58pt}}
&3&\multirow{4}{0.5pt}[4.8pt]{\rule{0.5pt}{58pt}}\\
\hline
&1&&$a$&&$d$&&$e$&\\
\hline
&2&&$d$&&$b$&&$f$&\\
\hline
&3&&$e$&&$f$&&$c$&\\
\hline
\end{tabular}
\]
where the row and column numbers will be used later to define $\mathrm{TP}_2$
inequalities. There are ${{3}\choose{2}}\times{{3}\choose{2}}$ or nine
possible combinations of four cells---that is, nine inequalities.
However, by
symmetry, three are simply duplicates of others. The following tableau
represents all of the inequalities:
\[
\begin{tabular}{@{}llccccccc@{}}
\hline\\[-10pt]
\multirow{4}{0.5pt}[4.8pt]{\rule{0.5pt}{58pt}}&&\multirow{4}{0.5pt}[4.8pt]{\rule{0.5pt}{58pt}}
&$(1,2)$$\vphantom{0^{0^0}}$&\multirow{4}{0.5pt}[4.8pt]{\rule{0.5pt}{58pt}}&$(1,3)$
&\multirow{4}{0.5pt}[4.8pt]{\rule{0.5pt}{58pt}}&$(2,3)$
&\multirow{4}{0.5pt}[4.8pt]{\rule{0.5pt}{58pt}}\\
\hline
&$(1,2)$&&$\mathbf{ab}\ge \mathbf{d}^2$&&$af\ge de$&&$df\ge be$&\\
\hline
&$(1,3)$&&$af\ge de$&&$ac\ge e^2$&&$dc\ge ef$&\\
\hline
&$(2,3)$&&$\mathbf{df}\ge \mathbf{be}$&&$dc\ge ef$&&$\mathbf{bc}\ge \mathbf{f}^2$&\\
\hline
\end{tabular}
\]
where $(i,j)\times(\ell,m)$ means form a matrix with elements from
rows $i$
and $j$ and columns $\ell$ and $m$ of the first tableau. Observe that when
the three inequalities highlighted in bold are satisfied, all others
will be
also satisfied. In fact, the inequality $(1,3)\times(1,2)\dvtx af\ge de$ derives
from $(1,2)\times(1,2)\dvtx\mathbf{ab}\ge\mathbf{d}^2$ and $(2,3) \times(1,2)\dvtx\mathbf{df}\ge\mathbf{be}$.
Finally, inequality $(2,3)\times(1,3)\dvtx dc\ge ef$ derives from the other two,
previously established---viz., $(2,3)\times(1,2)\dvtx\mathbf{df}\ge\mathbf{be}$ and
$(2,3)\times(2,3)\dvtx bc\ge f^2$. All other inequalities can be obtained from
the adjacent ones in this fashion.

Adding values to the type spaces of bidders expands the number of determinental
restrictions required to satisfy $\tptwo$, thus restricting the space of
distributions that can be entertained. Likewise, adding bidders to the game,
particularly if the bidders are assumed symmetric, also restricts the space
of distributions that can be entertained. For example, suppose a third bidder
is added, one who is symmetric to the previous two. The probability mass
function for triplets of values $(v_1,v_2,v_3)$, where $v_n=1,2,3$ and
$n=1,2,3$, can be represented as an array whose slices can then be represented
by the following three matrices for bidders $1$ and $2$, indexed by the values
of bidder~$3$:
\[
\bP_1=
\pmatrix{
a&d&e\cr
d&b&f\cr
e&f&c},\qquad
\bP_2=
\pmatrix{
d&b&f\cr
b&h&g\cr
f&g&i}\quad\mbox{and}\quad
\bP_3=
\pmatrix{
e&f&c\cr
f&g&i\cr
c&i&j}.
\]
In general, if the number of bidders is $N$ and the number of values is $k$,
then, without symmetry or affiliation, probability arrays have
$(k^N-1)$ unique
elements. Also, \citet{decastro2008} has shown that symmetry reduces
this to
${{k+N-1}\choose{k-1}}$ elements, while affiliation restricts where these
${{k+N-1}\choose{k-1}}$ probabilities can live on the simplex via the
determinental inequalities required by $\tptwo$. It is well known that a
function is $\mtptwo$ (affiliated), if and only if, it is $\tptwo$ in all
relevant collections of four points. As an aside, in this three-by-three
example, only nine constraints are relevant---viz.,
\begin{eqnarray*}
ab&\ge&d^2,\qquad bc\ge f^2,\qquad df\ge be,\qquad dh\ge b^2,\qquad hi\ge g^2,\\
bg&\ge&fh,\qquad eg\ge f^2,\qquad gj\ge i^2\quad\mbox{and}\quad fi\ge cg.
\end{eqnarray*}
If these hold, then the remainder are satisfied, too. Knowing the maximum
number of binding constraints is relevant later in the paper when we discuss
our test statistic.

Consider now the random $N$-vector $\bmV$ which equals $(V_1,\ldots,V_N)$,
having joint density (mass) function $f_{\bmV}$ with realization
$\mathbf{v}$ equal
to $(v_1,\ldots,v_N)$. Affiliation can be formally defined as follows: for
all $\mathbf{v}$ and $\mathbf{v}'$, the random variables $\bmV$ are
said to be affiliated if
\[
f_{\bmV}(\mathbf{v}\vee\mathbf{v}')f_{\bmV}(\mathbf{v}\wedge
\mathbf{v}') \ge
f_{\bmV}(\mathbf{v})f_{\bmV}(\mathbf{v}'),
\]
where
\[
(\mathbf{v}\vee\mathbf{v}') = [\max(v_1,v_1'),\max
(v_2,v_2'),\ldots,\max(v_N,v_N')]
\]
denotes the component-wise maxima of $\mathbf{v}$ and $\mathbf{v}'$,
sometimes referred
to as the \textit{join}, while
\[
(\mathbf{v}\wedge\mathbf{v}') = [\min(v_1,v_1'),\min
(v_2,v_2'),\ldots,\min(v_N,v_N')]
\]
denotes the component-wise minima, sometimes referred to as the \textit{meet}.

\section{Theoretical model}\label{s3}

We develop our theoretical model within the private-values paradigm, assuming
away any interdependencies. We consider a set $\mathcal N$ of bidders $\{1,2,
\ldots,N\}$. Now, bidder $n$ is assumed to draw $V_n$, his private valuation
of the object for sale, from the closed interval $[\underline
{v},\overline{v}]$.
We note that, without loss of generality, one can reparametrize the valuations
from $[\underline{v}, \overline{v}]$ to $[0,1]$. Below, we do this. We
collect the valuations in the vector $\mathbf{v}$ which equals
$(v_1,\ldots,v_N)$
and denote this vector without the $n$th element by $\mathbf{v}_{-n}$. Here,
we have used the now-standard convention that upper-case letters denote random
variables, while lower-case ones denote their corresponding realizations.
Note, too, that $\bmV$ lives in $[0,1]^N$.

We assume that the values are distributed according to the probability density
function $f_{\bmV}\dvtx[0,1]^N \rightarrow\Real_+$ which is symmetric;
that is, for
the permutation $\varphi\dvtx\{1,\ldots, N\} \rightarrow
\{1,
\ldots,N\},$ we have $f_{\bmV}(v_1,\ldots,v_N)$ equals
$f_{\bmV}(v_{\varphi(1)},\break \ldots,v_{\varphi(N)})$.
Letting $f_n(v_n)$ denote the marginal probability density function of $V_n$,
we note that it equals $\int_0^1\cdots\int_0^1 f_{\bmV}(\mathbf
{v}_{-n},v_n)\,\dd\mathbf{v}_{-n}$. [Below, we constrain ourselves to the case where
$f_n(\cdot)$
is the same for all $n$, but this is unnecessary and done only because,
when we
come to apply the method, we do have not enough information to estimate the
case with varying $f_n$'s.] Our main interest is the case when $f_{\bmV
}$ is
\textit{not} the product of its marginals---the case where the types are
dependent. We denote the conditional density of $\bmV_{-n}$ given
$v_n$ by
\[
f_{\bmV_{-n}|V_n}(\mathbf{v}_{-n}|v_n) = \frac{f_{\bmV}(\mathbf{v}_{-n},v_n)}{f_n(v_n)}.
\]
Finally, we denote the largest order statistic of $\bmV_{-n}$ given
$v_n$ by
$Z_n$ and its probability density and cumulative distribution functions by
$f(z_n|v_n)$ and $F(z_n|v_n)$, respectively.

We assume that bidders are risk neutral and abstract from a reserve price.
Given his value $v_n$, bidder $n$ tenders a bid $s_n\in\Real_+$. If his
tender is the highest, then bidder $n$ wins the object and pays what he
bid. A
pure strategy is a function $\sigma\dvtx[0,1]\rightarrow\Real_+$ which specifies
the bid $\sigma(v_n)$ for each value $v_n$. The interim pay-off
of bidder $n$, who bid $s_n$ when his opponents follow $\sigma\dvtx[0,1]
\rightarrow\Real_+$, is
\[
\Pi(v_n,s_n,\sigma)= (v_n-s_n)\int_{\underline{v}}^{\sigma^{-1}(s_n)}
f(z_n|v_n)\,\dd z_n
= (v_n-s_n)F[\sigma^{-1}(s_n)|v_n].
\]
We focus on symmetric, increasing pure-strategy equilibria (PSE) which are
defined by $\sigma\dvtx[0,1]\rightarrow\Real_+$ such that
\[
\Pi[v_n,\sigma(v_n),\bsigma_{-n}]\geq
\Pi(v_n,s,\bsigma_{-n})\qquad\forall s,v_n.
\]

As mentioned above, in most theoretical models of auctions that admit
dependence in valuation draws, researchers have assumed that $f_{\bmV}$
satisfies affiliation. We do not restrict ourselves to $f_{\bmV}$'s that
satisfy affiliation. We assume only that $f_{\bmV}$ belongs to a set of
distributions $\cP$ which guarantees the existence and uniqueness of a MPSE.
This set $\cP$ was fully characterized by \citet{decastro2008} in the
particular case of grid distributions, which are considered in our
Assumption \ref{ass:grid}, below.

Let $\cC$ denote the set of continuous density functions
$f_{\bmV}\dvtx[0,1]^N\rightarrow\Real_+$ and let $\cA$ denote the set
of affiliated
probability functions. For convenience and consistency with the
notation used
in later sections, we include in $\cA$ the set of all affiliated probability
functions, not just the continuous ones. Endow $\cC$ with the topology
of the
uniform convergence---that is, the topology defined by the norm of the supremum
\[
\Vert f_{\bmV}\Vert=\sup_{\mathbf{v}\in[0,1]^N}\vert f_{\bmV}(\mathbf{v})\vert.
\]
Let $\cD$ be the set of probability functions $f_{\bmV}\dvtx[0,1]^N\rightarrow
\Real_+$ and assume there is a measure $\mu$ over it.

We now introduce a transformation $\bbT^k\dvtx\cD\rightarrow\cD$ which
is the
workhorse of our method. To define $\bbT^k$, let $\Int\dvtx[0,1]\rightarrow
\{1,2,\ldots,k\}$ denote the function that associates to $v\in[0,1]$ the
ceiling $\lceil kv\rceil$---viz., the smallest integer at least as
large as
$kv$. Thus, for each $v\in[0,1]$, we have $v\in(\frac{\Int
(v)-1}{k},\frac{\Int(v)}{k}]$. Similarly, let $\bbS(\mathbf{v})$
denote the ``square''
(hypercube) $\prod_{n=1}^N(\frac{\Int(v_n)-1}{k},\frac{\Int
(v_n)}{k}
]$ where $\mathbf{v}$ collects $(v_1,v_2,\ldots,v_N)\in
[0,1]^N$. From this,
we define $\bbT^k\dvtx\cD\rightarrow\cD$ as the transformation that
associates to
each $f_{\bmV}\in\cD$ the probability density function $\bbT
^k(f_{\bmV})$ given
by
\[
\bbT^k(f_{\bmV})(\mathbf{v}) = k^N\int_{\bbS(\mathbf{v})} f_{\bmV
}(\bmu)\,\dd\bmu.
\]
Observe that $\bbT^k(f_{\bmV})$ is constant over each square
$\prod_{n=1}^N(\frac{m_n-1}{k},\frac{m_n}{k}]$, for all
combinations of $m_n\in\{1,\ldots,k\}$. The term $k^N$ above derives
from the
fact that each square $\prod_{n=1}^N(\frac{m_n-1}{k},
\frac{m_n}{k}]$ has volume $(1/k^N)$. Note that for all probability
density functions $f_{\bmV}\in\cD$, $\bbT^1(f_{\bmV})(\mathbf
{v})$ equals one for all
$\mathbf{v}\in[0,1]^N$, that is, $\bbT^1(f_{\bmV})$ is the uniform
distribution on
$[0,1]^N$.

We now need to introduce a compact notation to represent arrays of dimension
$\overbrace{k\times k\times\cdots\times k}^{N\ \mathrm{times}}$. We
denote by
$\cM^{k^N}$ the set of arrays and by $[\bP]$ an array in that set. When
there are but two bidders, an array is obviously just a matrix. In the
application of this model to field data, which we describe in Section
\ref{s5}, $N$
is three. The $(i_1,i_2,\ldots,i_N)$th element of an array is denoted
$[\bP](i_1,i_2,\ldots,i_N)$, or $[\bP](\bmi)$ for short, where
$\bmi$ denotes
the vector $(i_1,i_2,\ldots,i_N)$. Now, $\Int(v)=i$ if
$v\in(\frac{i-1}{k},\frac{i}{k}]$. Thus, for $k\in
\Nat$,
we define the finite-dimensional subspace $\cD^k\subset\cD$ as
\[
\cD^k=\{f_{\bmV}\in\cD\dvtx\exists[\bP]\in\cM^{k^N},\
f_{\bmV}(\mathbf{v}) = [\bP][\Int(v_1),\ldots,\Int(v_N)]\}.
\]
Observe that $\cD^k$ is a finite-dimensional set. In fact, when $N$ is two,
a probability density function $f_{\bmV}\in\cD^k$ can be described
by a
$(k\times k)$ matrix $\bP$ as follows:
\begin{equation}\label{deffn}
f_{\bmV}(v_1,v_2)=[\bP](i,j)\qquad
\mbox{if }(v_1,v_2)\in\biggl(\frac{i-1}{k},\frac{i}{k}\biggr]
\times
\biggl(\frac{j-1}{k},\frac{j}{k}\biggr]
\end{equation}
for $i,j\in\{1,2,\ldots,k\}.$ The definition of $f_{\bmV}$ at the
zero measure
set of points $\{(v_1,v_2)=(\frac{i}{k},\frac{j}{k})\dvtx i=0\mbox{ or }j=0\}$
is arbitrary.

\begin{figure}

\includegraphics{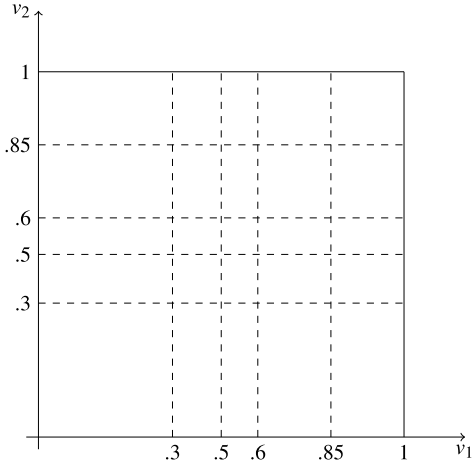}

\caption{Symmetric nonequi-spaced grid.}\label{fig:31}
\end{figure}

Note, too, that the width of the cells can be allowed to vary. For example,
one might be $0.3$ wide, while the next one can be $0.2$ wide, the third
$0.1$ wide, the next $0.25$, and the last $0.15$. In fact, the
transformation can be defined in terms of rectangles, instead of
squares as
above. To illustrate this, consider again the symmetric case and introduce
Figure \ref{fig:31}. Let $0=r_0<r_1<r_2<\cdots<r_{k-1}<r_k=1$ be an
arbitrary partitioning of the interval $[0,1]$.\footnote{We implicitly assume here that the $r_1<\cdots<r_k$ that form
$\bbB$ become dense in $[0,1]$ as $k$ increases.}
Now, define $\bbI\dvtx[0,1]\rightarrow\{1,2,\ldots,k\}$ by $\bbI(v)=j$
if and
only if $v\in(r_{j-1},r_j]$. Define $\bbB(\mathbf{v})$ as the
rectangle (box) where
$\mathbf{v}$ collects $(v_1,\ldots,v_N)\in[0,1]^N$ lies. Thus, $\bbB
(\mathbf{v})\equiv
\prod_{n=1}^N(r_{\bbI(v_n)-1},r_{\bbI(v_n)}]$. Now, define
\[
\bbT_{\bbB}^k(f_{\bmV}^0)(\mathbf{v}) =
\frac{\int_{\bbB(\mathbf{v})}f_{\bmV}^0(\bmu)\,\dd\bmu}{\int_{\bbB(\mathbf{v})}\,\dd\bmu}.
\]
The following theorem was proven by \citet{decastro2008}:

\begin{Thm}
Let $f_{\bmV}^0$ be a symmetric and continuous probability density function.
$f_{\bmV}^0$ is affiliated if and only if for all $k$,
$\bbT_{\bbB}^k(f_{\bmV}^0)$ is also affiliated.
\label{thm:31}
\end{Thm}

In our notation,
\[
f_{\bmV}^0\in\cA\quad\Leftrightarrow\quad\bbT_{\bbB}^k(f_{\bmV}^0)\in\cA
\qquad\forall k\in\bbN
\]
or
\[
\cA=\bigcap_{k\in\bbN}\bbT_{\bbB}^{-k}(\cA\cap\cD^k).
\]
Why is this important? Well, in many applications, the set of hypercubes
defined by a large $k$ will have many empty cells, which causes
problems in
both estimation and inference. Thus, one may want to subdivide the
space of
valuations irregularly, but symmetrically, as illustrated in Figure
\ref{fig:31} when $N$ is two.

\section{Test of affiliation}\label{s4}

The key result from \citet{decastro2007} that allows us to develop
our test of
symmetric affiliation is the following: if the true probability density
function $f_{\bmV}^0$ exhibits affiliation, then $\bbT_\bbB
^k(f_{\bmV}^0)$, a
discretized version of it, will too. (See Theorem \ref{thm:31},
above.) To
the extent that the grid distribution $\bbT_\bbB^k(f_{\bmV}^0)$ can be
consistently estimated from sample data, one can then test whether the
estimated grid distribution exhibits affiliation. Of course, sampling error
will exist, but presumably one can evaluate its relative importance using
first-order asymptotic methods.

Consider a sequence of $T$ auctions indexed $t=1,\ldots,T$ at which
$N$ bidders
participated by submitting the $NT$ bids $\{\{s_{nt}\}_{n=1}^N\}_{t=1}^T$.
We note that affiliation is preserved under a monotonic transformation, so
examining a discretization of $g_{\bmS}^0(\mathbf{s})$, the true
probability density
function of bids under the hypothesis of expected-profit maximizing equilibrium
behavior, is the same as examining $f_{\bmV}^0(\mathbf{v})$. Of
course, neither
$f_{\bmV}^0$ nor $g_{\bmS}^0$ is known. One can, however, construct an
estimate of $\bbT_\bbB^k(g_{\bmS}^0)$ on the interval $[0,1]^N$ by first
transforming the observed bids according to
\[
u_{nt} = \frac{s_{nt}-\lows}{\upps-\lows},\qquad n=1,\ldots,N\mbox{ and } t=1,\ldots,T,
\]
where $\lows$ is the smallest observed bid and $\upps$ is the largest observed
bid, and then by breaking up this hypercube into $L(=k^N)$ cells and counting
the number of times that a particular $N$-tuple falls in that cell.\footnote{We know that the support of $g_{\bmS}^0$ is strictly
positive at
$\sigma^0(\overline{v})$, the true upper bound of support of bids,
and we assume that $f_{\bmV}^0$ is strictly positive at
$\underline{v}$, so $g_{\bmS}^0$ is strictly positive at
$\sigma^0(\underline{v})$, the true lower bound of support of bids.
Consequently, the sample estimators of the lower and upper bounds
of support of $\bmS$ converge at rate $T$, which is faster than the
rate of convergence of sample averages---rate $\sqrt{T}$. Hence,
when using sample averages in our estimation, we can ignore this
first-stage, pre-estimation error---at least under first-order
asymptotic analysis.}
Now, the random vector $\bmY$, which represents the number of outcomes that
fall in each of the cells and equals the vector $(Y_1,Y_2,\ldots
,Y_L)^\top$,
follows a multinomial distribution having the joint probability mass function
\[
g_{\bmY}(\mathbf{y}|\bolds{\pi}) =
\frac{T!}{y_1!\cdots y_L!}\prod_{\ell=1}^L\pi_\ell^{y_\ell},
\]
where $\pi_\ell$ equals $\Prr(Y_\ell=y_\ell)$, with $y_\ell
=0,1,\ldots,T$, while
$\bolds{\pi}$ collects $(\pi_1,\ldots,\pi_L)$ and lives on the
simplex---viz., the set
\[
\cS_L = \{\bolds{\pi}|\bolds{\pi}\ge\bzero_L,\biota_L^\top
\bolds{\pi}=1\}
\]
with $\biota_L$ being an $(L\times1)$ vector of ones. Note, too, that
$\biota^\top\mathbf{y}$ equals $T$, the number of observations.

For $\ell=1,\ldots,L$, the unconstrained maximum-likelihood estimates of
the $\pi_\ell$'s are the $(y_\ell/T)$'s. To test for affiliation, maximize
the following logarithm of the likelihood function (minus a constant):
\[
\cL(\bolds{\pi}) = \mathbf{y}^\top\log(\bolds{\pi})
\]
subject to
\begin{longlist}
\item[(1)] the vector $\bolds{\pi}$ lies in the simplex $\cS_L$;
\item[(2)] all of the determinental inequalities required for $\mathrm{TP}_2$
hold.
\end{longlist}
Then compare this value of $\cL$ with the unconstrained one.

While the determinental constraints required for $\tptwo$ are convex
sets of
the parameters when the submatrices are symmetric, they are not for general
submatrices. However, by taking logarithms of both sides of any general
determinental inequality
\[
ab\ge cd,
\]
one can convert this into a linear inequality, which does give rise to
convex constraint sets, albeit in variables that are logarithms of the original
variables. To wit,
\[
\log a+ \log b - \log c -\log d\ge0
\]
defines a convex set (in the transformed variables $\log a,\ldots,\log d$).
Of course, the adding-up constraint for the simplex must be
finessed---for example, by
considering the following:
\[
\exp(\log a)+\exp(\log b)+\exp(\log c)+\exp(\log d)+\cdots\le1,
\]
which gives rise to a convex set. Thus, the problem is almost a linear
programme.

For known $N$ and fixed $k$, the specific steps involved in
implementing the
test in this problem are the following. First, form the grid
distribution of
the joint density as the unknown array $[\bP]$. Letting $[\bE]$
denote the
array of counts for the grid distribution, the logarithm of the likelihood
function for this multinomial process is
\begin{equation}\label{logL}
\sum_{\bmi} [\bE](\bmi)\log\{[\bP](\bmi)\}.
\end{equation}
Now, the following inequalities must be met:
\begin{equation}\label{simplex}
\log\{[\bP](\bmi)\}\le0\quad\mbox{and}\quad
\sum_{\bmi}\exp(\log\{[\bP](\bmi)\})\le1,
\end{equation}
while symmetry requires the following linear restrictions:
\begin{equation}\label{symmetry}
[\bP](\bmi) = [\bP][\varphi(\bmi)]
\end{equation}
where $\varphi(\cdot)$ is any permutation, and affiliation requires the
following determinental inequalities:
\begin{equation}\label{affil}
\log\biggl\{\frac{[\bP](\bmi\vee\bmi')[\bP](\bmi\wedge\bmi')}{[\bP](\bmi)[\bP](\bmi')}\biggr\} \ge0
\end{equation}
hold. A test of affiliation, within a symmetric environment, involves
comparing the maximum of equation (\ref{logL}), subject to the
constraints in
(\ref{simplex}) and (\ref{symmetry}), with the maximum of equation
(\ref{logL}), subject to the constraints in (\ref{simplex}), (\ref{symmetry})
and (\ref{affil}).

Our test of symmetric affiliation is based on the difference between the
maximum of the logarithm of the likelihood function $\cL([\hat\bP])$
and the
maximum of the logarithm of the likelihood function under symmetric affiliation
$\cL([\tilde\bP])$. Obviously, the sampling theory associated with the
difference in these two values of the objective function $\cL$ is not
straightforward because not all of the inequality constraints required by
$\mtptwo$ may hold and, from sample to sample, the ones that do hold can
change, but we shall suggest several strategies to deal with this, below.

Experience gleaned from other models with a related structure---for
example, Wolak (\citeyear{wolak1987}, \citeyear{wolak1989b},
\citeyear{wolak1989a}, \citeyear{wolak1991}) as well as
\citet{bartforc2000}, who investigated $\mtptwo$ in binary models---suggests
that the likelihood ratio (LR) statistic
\begin{equation}\label{lrstat}
2[\cL([\hat\bP])-\cL([\tilde\bP])]
\end{equation}
is not distributed according to a standard $\chi^2$ random variable.

Introducing $\vecc[\bP]$ as a short-hand notation, for the $L$-vector created
from the array $[\bP]$, our constrained-optimization problem can be summarized
in a notation similar to that of Wolak (\citeyear{wolak1989a}) as
\[
\max_{\vecc[\bP]}\mathbf{y}^\top\log(\vecc[\bP])\quad\mbox
{subject to}\quad
\mathbf{h}(\vecc[\bP])\geq\bzero_J,
\]
where $\mathbf{h}\dvtx\Real^L\rightarrow\Real^J$ is the function
representing all $J$
relevant constraints where $J\leq L$ and $L$ is the total number of variables
under the alternative hypothesis. (Here, for notational parsimony, we have
ignored the adding-up condition, which is implicit.)

Consider $N_{\bdelta}(\vecc[\bP^0])$, a neighborhood of the true value
$\vecc[\bP^0]$. Denote by $\bH(\vecc[\bP^0])$ the matrix of partial
derivatives
whose $(i,j)$-element is $\frac{\partial h_i(\vecc[\bP])}{\partial
\vecc[\bP]_j}$. Now, let us define the set
$\cB=\{\vecc[\bP]\dvtx\bH(\vecc[\bP^0])\vecc[\bP] \geq\bzero,\vecc
[\bP]\in\Real^L\}$.
Denote by $\bcI(\vecc[\bP^0])$ Fisher's information matrix which is
defined by
\[
\lim_{T\rightarrow\infty} T^{-1}\cE_{[\bP^0]}
\biggl[-\frac{\partial^2\cL(\vecc[\bP])}{\partial\vecc[\bP]\,\partial\vecc[\bP]^\top}\biggr]
\]
evaluated at $\vecc[\bP^0]$. Finally, denote by
\[
\bPi^0=\bH(\vecc[\bP^0])\bcI(\vecc[\bP^0])^{-1}\bH(\vecc[\bP
^0])^\top
\]
the variance--covariance matrix of $\mathbf{h}(\vecc[\hat\bP])$ and by
$\omega(j,J-j,\bPi^0)$, the probability that $j$ constraints bind, that
$(J-j)$ constraints are strictly satisfied, that is, they are
nonbinding. We
have the following:

\begin{Thm}\label{thm:41}
Consider the local hypothesis testing problem
\begin{eqnarray*}
\mathrm{H}_0&\dvtx&\mathbf{h}(\vecc[\bP])\geq\bzero_J\vecc[\bP]\in N_{\bdelta}(\vecc[\bP^0]),\\
\mathrm{H}_1&\dvtx&\mbox{not }\mathrm{H}_0.
\end{eqnarray*}
The asymptotic distribution of the likelihood-ratio statistic satisfies the
following property:
\[
\sup_{b\in\cB}\
\Prr_{[\bP^0],\bcI(\vecc[\bP^0])^{-1}}(D\geq c) =
\Prr_{[\bP^0]}(D\geq c) =
\sum_{j=0}^J\Prr(W_j\geq c)\omega(j,J-j,\bPi^0),
\]
where $D$ is the asymptotic value of the test statistic, while $W_j$ is an
independent $\chi^2$ random variable having $j$ degrees of freedom.
\end{Thm}

\begin{pf}
It is sufficient and straightforward to verify that the assumptions of Theorem
4.2 in \citet{wolak1989b} are satisfied.
\end{pf}
Because this statistic depends on the unknown population grid
distribution $[\bP^0]$, the statistic is not pivotal.
\citet{koddpalm1986} have provided lower and upper bounds for this test
statistic for tests of various sizes and different numbers of maximal
constraints.

According to \citet{wolak1989b}, the best way to evaluate the weights
is using Monte Carlo simulation. Wolak also offered lower and upper
bounds for the probabilities above [see his equations (19) and (20),
page 215]; these bounds are based on \citet{koddpalm1986}. An
alternative strategy would be to adapt the bootstrap methods of
\citet{bugni2008} to get the appropriate \textit{p}-values of the test
statistic. Yet a third strategy would be to adapt the subsampling
methods described in \citet{polietal1999} as was done by
\citet{romashai2008}.

\subsection{Some comparisons with other nonparametric methods}\label{s41}

It should be noted, too, that our proposed estimation strategy involves
nothing more than estimating a histogram using a special class of
grids. Scott [(\citeyear{scott1992}), page xi] has argued that the
classical histogram ``remains the most widely applied and most
intuitive nonparametric estimator.'' In other words, the procedure
proposed here is not based on any unfamiliar concepts. Of course, there
are more statistically efficient methods, but they also have
limitations, as \citet{scott1992} has discussed. Also, although the
rate of convergence of histogram estimation is slow, it is still
reasonable; see Scott [(\citeyear{scott1992}), Theorem 3.5, page 82].

Note, too, the similarities between grid-distribution and
kernel-smoothed estimators. Kernel-smoothed density estimators are
well-behaved and have good rates of convergence when the probability
density functions to be estimated are continuously differentiable
$\C^1$.\footnote{Methods exist that require fewer smoothness
conditions---for example, the function need just be continuous $\C^0$;
others require additional smoothness, $\C^2$ or higher. This does not
change our claims.} The set $\C^1$ is dense in the set of all
probability density functions. Similarly, grid distribution estimators
are well-behaved for probability density functions in
$\cD^{\infty}=\bigcup_{k=1}^\infty\cD^k$, which is also a dense set in
the set of all probability density functions.\footnote{Recall that
$\cD^k$ is the set of grid distributions where the interval is
subdivided into $k$ intervals, that is, $\cD^k\equiv \bbT^k(\cD).$}
While $\C^1$ probability density functions form a familiar and
well-known class probability density functions, the probability density
functions in $\cD^k$ are also familiar because they are just (a special
class of) simple functions, which are fundamental, such as in the
definition of the Lebesgue integral. When estimating grid
distributions, one has to choose $k$ or, equivalently, the size of the
bin $(1/k)$, which is nothing more than the bandwidth of the
grid-distribution estimator. Similarly, kernel-smoothing requires a
choice of bandwidth parameter, too. In sum, nonparametric estimation
using either grid distributions or smoothed kernels is very similar.

\subsection{Consistency and power of the proposed test}\label{s42}

Of course, one concern is that $k$ appears fixed in our analysis, but $T$
is increasing, so our test is potentially inconsistent. We imagine a sequence
of $\{k_T\}$ with values increasing as $T$ increases, but not as fast
as $T$.
Below, we discuss in detail what we have in mind. Another worry is that the
test statistic will be ill-behaved if $k_T$ tends to infinity. Thus, an upper
bound $\bar k$ must exist. This discussion leads us to introduce the following
assumption concerning $f^0_{\bmV}$ which allows us to side-step these technical
problems:

\begin{ass}\label{ass:grid}
The true data-generating process $f^0_{\bmV}$ is a grid distribution,
that is, there exists $\bar k\in\mathbb{N}$ such that
$f^0_{\bmV}\in\cD^{\bar k}$.
\end{ass}

As the discussion above made clear, this assumption is similar to the
assumptions of smoothness concerning $f^0_{\bmV}$ which kernel-smoothing
methods require. In addition to this analogy, we offer two additional
justifications for Assumption \ref{ass:grid}.

First, the set of grid distributions is dense in the set of all distributions:
even if the data-generating process (DGP) $f^0_{\bmV}$ were not a grid
distribution, there is a grid distribution that is arbitrarily close to it.
To wit, no finite amount of data could reject Assumption \ref
{ass:grid}. In
this sense, Assumption \ref{ass:grid} is almost ``no assumption.''

Second, the DGP in question is a distribution of values, which are discrete
(up to, say, dollars or cents or Yen or Won or whatever units one wants).
When one assumes a smooth probability density function, one is making an
approximation, for computational convenience: such an approximation does
not seem, to us at least, any more appealing than the one we make. On the
contrary, it seems more natural to us to assume simple probability density
functions rather than any smoothness conditions. In general, smoothness is
just a tool used to lighten the burden in the technical analysis of a
particular problem. In our case, by assuming that the distribution is simple
(i.e., a grid distribution), we can stay closer to reality.

Under Assumption \ref{ass:grid}, our test is consistent, for Assumption
\ref{ass:grid} implies that a $k$ exists such that $f_{\bmV}^0\in\cD^k$.
Therefore, the number of inequalities required for affiliation remains
fixed. We are then in the standard framework considered by Wolak, which
has a fixed set of inequalities. Thus, consistency follows directly from
Wolak's research. A technically sophisticated reader may feel that our
consistency result is trivial, once Assumption \ref{ass:grid} is made. The
point of this paper (and this subsection, in particular) is not to provide
a technical proof of consistency, but rather to remove any doubts concerning
the consistency of our test under a reasonable assumption.

For any specific implementation, $k$ is assumed fixed in the
approximation. In
the asymptotics, we imagine $k_T$ increasing as $T$ increases, until some
upper bound $K$ is reached. In any application, however, if $T$ is quite
large, not what we encounter in our application, then one can vary $k$, which
will potentially yield different estimates.

The power of the proposed test clearly depends on the choice of $k$.
Were $k$
chosen to be one (i.e., a uniform distribution on the $N$-dimensional
hypercube), then affiliation would never be rejected. On the other hand,
given a finite sample of $T$ observations, a large $k$ will result in many
cells having no elements. While the choice of $k$ is obviously important
and certainly warrants additional theoretical investigation, perhaps along
the lines of research in time-series analysis by \citet{guayetal2008}
concerning optimal adaptive detection of correlation functions, it is beyond
the scope of this paper. In fact, in most applications to auctions, where
samples are often quite small, $k$ will be dictated by practical
considerations---viz., the relative size of $T$.

\subsection{Bounding the number of inequalities}\label{s43}

For our test statistic to be well-behaved, it is important to know that
an upper bounds exists on the number of inequalities. For arbitrary $N$
and $k$, assuming a symmetric distribution, we can construct a bound on
how many inequalities there are. Because we focus on symmetric
distributions,
\[
[f^0_{\bmV}](i_1,i_2,\ldots,i_N)=
[f^0_{\bmV}](i_1^\prime,i_2^\prime,\ldots
,i_N^\prime),
\]
where $(i_1^\prime,i_2^\prime,\ldots,i_N^\prime)$ is a permutation of
$(i_1,i_2,\ldots,i_N)$. Thus, we need only consider sorted indices,
indices $(i_1,i_2,\ldots,i_N)$ for which $i_1\geq i_2 \geq\cdots\geq
i_N$. Consider $(i_1,i_2,\ldots,i_N)$, a sorted index having $\ell$
different numbers; let $r_1,\ldots,r_\ell$ denote the number of
repetitions of the different numbers in $(i_1,i_2,\ldots,i_N)$.
Obviously, $r_1+\cdots+r_\ell =N$. Using this notation, the number of
permutations of $(i_1,i_2,\ldots ,i_N)$ is then $\frac{N!}{r_1!\cdots
r_\ell!}$. For instance, the index $(4,3,3,2,2,2)$ has
$\frac{6!}{1!2!3!}$ or $120$ different permutations.

Given the above, we can now focus our attention to sorted indices only.
Consider the lexicographic order of them. In this way, we can attribute an
unambiguous natural number to each sorted index of length $N$. For example,
consider $N=3$, in which case $(1,1,1)$ corresponds to 1; $(2,1,1)$, to 2; $(2,2,1)$,
to 3; $(3,1,1)$ to 5; $(3,2,2)$, to 7; and $(4,1,1)$ to 11. It is important to
develop an algorithm to convert a sorted index into a corresponding number,
which we describe now.

First, let us define $\Num(j,N)$ as the number of all indices
that are\break weakly below (in the lexicographic order) to the index
$(j,j,\ldots,
j)$, that is, the index that has $j$ in all positions and has length
$N$. It is
easy to see\break that $\Num(1,N)=1$, because there is just one index weakly
below $(1,1,1,\break\ldots,1)\dvtx (1,1,1,\ldots,1)$, itself. Also, $\Num(2,2)=3$,
because $(1,1)$, $(2,1)$ and $(2,2)$ are the sorted indices weakly below
$(2,2)$. Similarly, $\Num(2,3)=4$, because $(1,1,1)$, $(2,1,1)$, $(2,2,1)$,
$(2,2,2)$ are the sorted indices weakly below $(2,2,2)$. From this argument,
it is not difficult to see that $\Num(2,N)=N+1$. Observe, too, that
$\Num(j,1)=j$, because there are only the indexes $(1)$, $(2)$,
$\ldots$\,, $(j)$
weakly below $(j)$. \citet{decastro2008} has proven the following:

\begin{lem}
$\Num(j,N)={{N+j-1}\choose{j-1}}$.
\end{lem}

Thus, if we fix the number of bidders $N$ and the number of intervals $k$,
then there are $M\equiv\Num(k,N)={{N+k-1}\choose{k-1}}$
different
indices. Affiliation will be satisfied if the corresponding inequality is
satisfied for any pair of indices $(\bmi,\bmi')$. Since there are
${{M}\choose{2}}$ or $\frac{M(M-1)}{2}$ pair of such indices, it is sufficient
to test $(M^2-M)/2$ inequalities. Note, however, that this is an upper
bound because some inequalities are implied by others. The above discussion
also provides some guidance concerning how to choose the inequalities; however,
in an effort to conserve space, we leave the discussion of what the minimal
set of sufficient inequalities is to another paper.

\subsection{Two related papers}\label{s44}

Like us, \citet{lizhan2008} have examined some important economic implications
of affiliation. Instead of considering bids, however, Li and Zhang
examined the
entry behavior of potential bidders whose signals may be affiliated. Theirs
is a parametric analysis and they implemented their test using simulation
methods, examining timber sales organized by the Department of Forestry
in the
State of Oregon. Li and Zhang found only a small degree of affiliation,
perhaps because the zero/one entry decision is not as informative as
bid data.

\citet{junetal2009} have developed a consistent nonparametric test designed
for continuous data. By avoiding discretization, Jun et~al.~presumably have
more information than we do. On the other hand, having rejected affiliation
with their test, it is unclear what to do within their framework
because an
alternative hypothesis is unspecified. In contrast, our approach
augments the
theoretical work of \citet{decastro2008} where the alternative hypothesis
is clearly outlined.

\subsection{Policy uses for grid distributions}\label{s45}

de Castro (\citeyear{decastro2008}) has developed a complete theoretical treatment of
grid distributions, even in the absence of
affiliation.\footnote{de~Castro's method is too long to be described in
detail here; his paper is more than seventy pages long. In a nutshell,
the method is as follows: first, it is shown that the usual proof of
uniqueness of monotonic pure-strategy equilibrium can be adapted to
grid distribution. Thus, if there is a monotonic pure-strategy
equilibrium, then it is unique and characterized by the solution to a
differential equation. Also, since we consider the symmetric case, an
explicit solution is available. In the case of grid distributions, this
solution is proven to be a rational function (a quotient of
polynomials). It is then shown that in each square defining a grid
distribution, it is sufficient to verify the equilibrium inequality
(optimality of following the bidding function) only with respect to a
finite number of pairs (types, bids). This step is necessary because, in
principle, one needs to check an infinite number of pairs (types, bids).
The final number of points to be tested is small (less than six) for
each square. Finally, it is proven that the candidate is an equilibrium
if and only if the test is satisfied. de Castro has also provided
expressions for revenues $R^1_{[\hat\bP]}$ when $[\hat\bP]$ is a grid
distribution.} His idea is as follows: first, assume that
$f^0_{\bmV}\in\cD^k$ for some $k$; that is, the DGP is a grid
distribution---Assumption \ref {ass:grid} holds. Standard estimation
methods (histograms) can be used to calculate $[\hat\bP]\in\cD^k$ that
best approximate $f^0_{\bmV}$.

Under de~Castro's method, one can then test whether $[\hat\bP]$ has a
symmetric MPSE. The method developed by de~Castro is exact: to wit, modulo
sampling error, $[\hat\bP]$ has a symmetric MPSE if and only if the method
detects it. Errors can occur only in simple numerical operations such
as sums,
divisions and square roots. It turns out that determining the existence
of a
symmetric MPSE is nontrivial when affiliation is absent.

If $[\hat\bP]$ has a symmetric MPSE, then it can be used to calculate
expected revenues under the first- and second-price auctions, denoted
$R^1_{[\hat\bP]}$ and $R^2_{[\hat\bP]}$, respectively. In this way, one
can determine which auction format yields a higher expected revenue for
$[\hat\bP]$ and, also, the magnitude of the revenue difference
$(R^2_{[\hat\bP]}-R^1_{[\hat \bP]})$, to decide whether it is
significant.\footnote{As explained above, de~Castro has shown that if a
monotone, pure-strategy equilibrium exists in a first-price auction,
then it is unique. Moreover, we can obtain the underlying distribution
of values from the distribution of bids, as is typically done in the
econometrics of auctions. Although second-price auctions may have
multiple equilibria, in general, in the literature, researchers
typically consider only the truthful bidding equilibrium. The
truth-telling equilibrium does not depend on the assumption of
affiliation: it is an equilibrium for any distribution. Thus, if we
have the distribution of values, we also have the distribution of bids
for this equilibrium.}

The procedure can then be repeated using $[\tilde\bP]$, which is
obtained under
the constraint that the distribution satisfies affiliation. We know that,
under affiliation, a~symmetric MPSE exists and that $(R^2_{[\tilde\bP]}-
R^1_{[\tilde\bP]})\geq0$, but the method also allows one to decide
whether the magnitudes of the differences $(R^1_{[\hat\bP
]}-R^1_{[\tilde\bP]})$
and $(R^2_{[\hat\bP]}-R^2_{[\tilde\bP]})$ are economically important.
It is quite possible that the expected-revenue difference between
first- and
second-price auctions is nonzero, but small in economic terms, and the method
allows one to examine sampling variability by repeating the above procedures
using resampled draws from $[\hat\bP]$ or $[\tilde\bP]$. Thus, if the
estimated difference is economically large relative to the sampling error,
then this is important information for a policy maker to know.

Thus, the grid distributions proposed in this paper have many advantages
because a theory exists that can be used for policy analysis. Such theories
have not yet been developed for other methods; if affiliation is rejected
under these methods, then what to do?

\section{Empirical application}\label{s5}
Above, in Section \ref{s3}, in the tradition of the theoretical
literature concerning auctions, we developed our model of bidding in
terms of valuations for an object to be sold at auction under the
first-price, sealed-bid \mbox{format}. Sealed-bid tenders are often used in
procurement---that is, low-price, sealed-bid auctions at which a buyer
(often a government agency) seeks to find the lowest-cost producer of
some good or service. In this section we report results from an
empirical investigation of procurement tenders for road resurfacing by
a government agency. Although it is well known that results from
auctions can be translated to procurement, and vice versa, sometimes
this translation is tedious. We direct the interested reader to the
work of \citet{decafrut2009} who have developed a procedure to
translate results from auctions to procurement.

We have applied our empirical framework to data from low-price,
sealed-bid, procurement auctions held by the Department of
Transportation (DOT) in the State of Michigan. At these auctions,
qualified firms are invited to bid on jobs that involve resurfacing
roads in Michigan. We have chosen this type of auction because the task
at hand is quite well understood. In addition, there are reasons to
believe that firm-specific characteristics make the private-cost
paradigm a reasonable assumption; for example, the reservation wages of
owners/managers, who typically are paid the residual, can vary
considerably across firms. On the other hand, other features suggest
that the cost signals of individual bidders could be dependent, perhaps
even affiliated; for example, these firms hire other factor services in
the same market and, thus, face the same costs for inputs such as
energy as well as paving inputs. For example, suppose paving at auction
$t$ has the following Leontief production function for bidder $n$:
\[
q_{nt} = \min\biggl(\frac{h_{nt}}{\delta_h}, \frac{y_{nt}}{\delta_y},
\frac{z_{nt}}{\delta_z}\biggr),
\]
where $h$ denotes the managerial labor, while $y$ and $z$ denote other
factor inputs which are priced competitively at $W_t$ and $X_t$,
respectively, at auction $t$. Assume that $R_n$, bidder $n$'s marginal
value of time, is an independent, private-cost draw from a common
distribution. In addition, assume that the other factor prices $W_t$
and $X_t$ are draws from another joint distribution, and that they are
independent of $R_n$. The marginal cost per mile $C_{nt}$ at auction
$t$ can be then written as
\[
C_{nt} = \delta_h R_n + \delta_y W_t + \delta_z X_t,
\]
which is a special case of an affiliated private-cost (APC) model,
known as a
\textit{conditional private-cost} model. The costs in this model are affiliated
only when the distribution of $R_n$ is log-concave, which is discussed
extensively in \citet{decastro2007}. \citet{lietal2000} have
studied this
model extensively. In short, the affiliated private-cost paradigm (APCP)
seems a reasonable null hypothesis.

We did not investigate issues relating to asymmetries across bidders because
we do not know bidder identities, data necessary to implement such a
specification. Because no reserve price exists at these auctions, we treat
the number of participants as if it were the number of potential
bidders and
focus on auctions at which three bidders participated. Thus, we are ignoring
the potential importance of participation costs which others, including
\citet{li2005}, have investigated elsewhere.

The data for this part of the paper were provided by the Michigan DOT
and were organized and used by \citet{hubbetal2009}; a complete
description of these data is provided in that paper. In Table
\ref{tab:51} we present the summary descriptive statistics concerning
our sample of $834$ observations---$278$ auctions that involved three
bidders each. We chose auctions with just three bidders not only to
illustrate the general nature of the method (if we can do three, then
we can do $N$), but also to reduce the data requirements. When we
subdivide the unit hypercube into $k^N$ cells, the average number of
bids in a cell is proportional to $(k^N/T)$. When $N$ is very large,
the sample size must be on the order of $k^N$ in order to expect at
least one observation in each cell. This example also illustrates the
potential limitations of our approach; viz., even in relatively large
samples, some of the cells will not be populated, so $k$ will need to
be kept small. However, one can circumvent this problem by varying the
width of the subdivisions as we do below. Of course, one must then
adjust the conditions which define the determinental inequalities. We
describe this below, too.

\begin{table}
\caption{Sample descriptive statistics---dollars$/$mile: $N = 3$;
$T=278$}\label{tab:51}
\begin{tabular}{@{}lccccc@{}}
\hline
\textbf{Variable} &\multicolumn{1}{c}{\textbf{Mean}} & \multicolumn{1}{c}{\textbf{St. Dev.}} & \multicolumn{1}{c}{\textbf{Median}} & \multicolumn{1}{c}{\textbf{Minimum}} &\multicolumn{1}{c@{}}{\textbf{Maximum}}\\
\hline Engineer's estimate &475,544.54 & 491,006.52& 307,331.26& 54,574.41& 3{,}694{,}272.59\\
Winning bid &466,468.63 & 507,025.81& 286,102.57& 41,760.32& 3,882,524.81\\
All tendered bids &507,332.42& 564,842.58& 317,814.77& 41,760.32& 5,693,872.81\\
\hline
\end{tabular}
\end{table}

Our bid variable is the price per mile. Notice that both the winning
bids as well as all tendered bids vary considerably, from a low of
\$41,760.32 per mile to a high of \$5,693,872.81 per mile. What
explains this variation? Well, presumably heterogeneity in the tasks
that need to be performed. One way to control for this heterogeneity
would be to retrieve each and every contract and then to construct
covariates from those contracts. Unfortunately, the State of Michigan
cannot provide us with this information, at least not any time soon.

How can we deal with this heterogeneity? Well, in our case, we have an
engineer's estimate $p$ of the price per mile to complete the
project.\footnote{Of course, besides $p$, it is possible that other
covariates, which are common knowledge to all the bidders, exist. If
these other common-knowledge covariates exist, then we could wrongly
conclude that the signals have a strong form of correlation when, in
fact, the correctly-specified model of signals (conditioned on the
common-knowledge information) would have only small correlation.
Unfortunately, we do not have access to any additional information.
Were such information available, then we would condition on it as
well.} We assume that $C_{nt}$, the cost to bidder $n$ at auction $t$,
can be factored as follows:
\begin{equation}\label{separ}
C_{nt} = \lambda^0(p_t)\varepsilon_{nt},
\end{equation}
where $\lambda^0$ is a known function. One example of this is
\[
C_{nt} = p_t\varepsilon_{nt}.
\]
Another is
\[
C_{nt} = \delta_0p_t^{\delta_1}\varepsilon_{nt}.
\]
Under equation (\ref{separ}), the equilibrium bid $B_{nt}$ at auction
$t$ for bidder $n$ takes the following form:
\[
B_{nt} = \lambda^0(p_t)\beta(\varepsilon_{nt}),
\]
so
\[
\frac{B_{nt}}{\lambda^0(p_t)} = \beta(\varepsilon_{nt}).
\]
Of course, we do not know $\lambda^0$, but we can estimate $\lambda ^0$
either parametrically, under an appropriate assumption, or
nonparametrically, using the following empirical specification:
\[
\log B_{nt} = \psi(p_t) + U_{nt},
\]
where $\psi(p_t)$ denotes $-\log[\lambda^0(p_t)]$ and $U_{nt}$ denotes
$\log[\beta(\varepsilon_{nt})]$.

\begin{figure}

\includegraphics{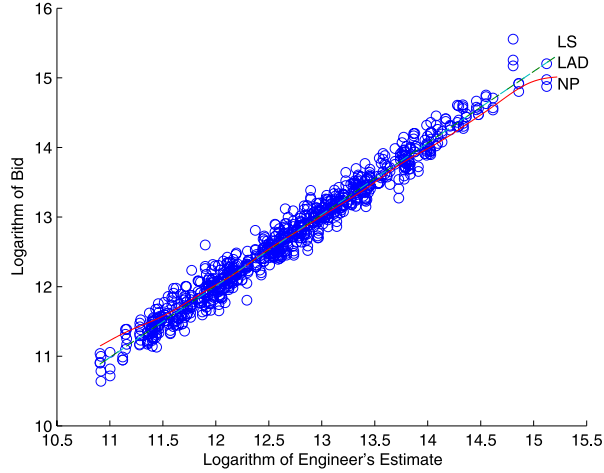}

\caption{Data as well as NP, LS and LAD regressions:
logarithm of bids versus logarithm of engineer's
estimate.}\label{fig:51}
\end{figure}

Empirical results from this exercise are presented in Figure
\ref{fig:51}. In this figure are presented results for the
nonparametric regression (NP), the least-squares regression (LS) and
the least-absolute-deviations (LAD) regression. To get some notion of
the relative fit, note that the $R^2$ for the LS regression is around
$0.97$. The LS estimates of the constant and slope coefficients are
$-0.3114$ and $1.0268$, respectively, while LAD estimates of the
constant and slope coefficients are $-0.3221$ and $1.0276$,
respectively.

\begin{figure}

\includegraphics{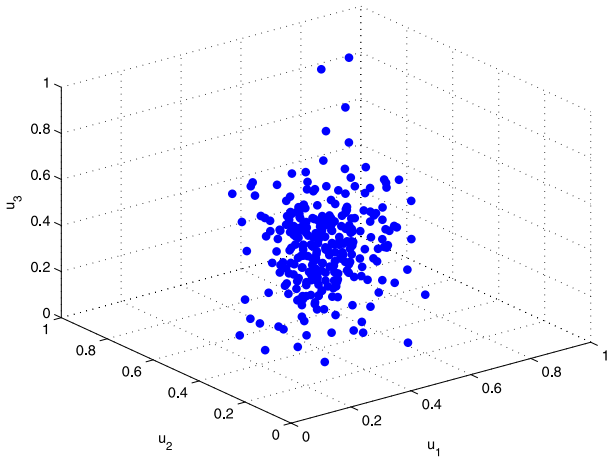}
 \caption{Scatterplot of transformed fitted LS
residuals.} \label{fig:52}
\end{figure}

Subsequently, we took the normalized fitted residuals, which (for the
LS case) are depicted in Figure \ref{fig:52}, and applied the methods
described in Section \ref{s4} above for a $k$ of two. Our test results
are as follows: the maximum of the logarithm of the likelihood function
(minus a constant) without symmetry was $-442.50$, while the maximum of
the logarithm of the likelihood function under symmetry was $-444.88$,
and under symmetric affiliation it was also $-444.88$---a total
difference of $2.38$.\footnote{The results for the LAD residuals were
identical: the probability array obtained by discretizing the LAD
residuals was exactly the same as in the LS case because none of the
fitted residuals was classified differently. This is not, perhaps,
surprising given the similar fits of the two empirical specifications.}
At size $0.05$, twice the above difference is above the lower bound
provided by \citet{koddpalm1986}, but below the upper bound, so the
test is inconclusive.

Because a $k$ of two is unusually small, we introduced a symmetric, but
nonequispaced, grid distribution---like the one depicted in Figure
\ref{fig:31}, but with intervals $[0,0.4)$, $[0.4,0.6)$ and
$[0.6,1.0]$. The $\tptwo$ inequalities can be derived in the usual way,
but the adding-up inequality must be rewritten, in this case as
\begin{eqnarray*}
&&a + 2b + 8c + 8d + 16e + 8f + 4d + 2h + 4i + 4b\\
&&\qquad{} + 16f + 8g + 8e + 2g + 8j + 8f + 16c + 8i \le1.
\end{eqnarray*}
Again, we applied our methods. Our test results are as follows: the
maximum of the logarithm of the likelihood function (minus a constant)
under symmetry was $-715.72$, while the maximum under symmetric
affiliation was $-716.49$---a difference of $0.77$.\footnote{Again, the
results for the LAD residuals were virtually identical: the probability
array obtained by discretizing the LAD residuals was almost the same as
in the LS case.} At size $0.05$, twice the above difference is below
the lower bound provided by Kodde and Palm, so we do not reject the
hypothesis of symmetric affiliation. To put these results into some
context, the center of the simplex had a logarithm of the likelihood
function of $-916.24$; using the marginal distribution of low, medium
and high costs $(0.4233,0.4808,0.0959),$ and imposing independence
yielded a logarithm of the likelihood function of $-784.67$.

\section{Summary and conclusions}\label{s6}

We have constructed a tractable empirical model of equilibrium behavior at
first-price auctions when bidders' private valuations are dependent,
but not
necessarily affiliated. Subsequently, we developed a test of
affiliation and
then investigated its small-sample properties. We applied our framework to
data from low-price, sealed-bid auctions used by the Michigan DOT to procure
road-resurfacing: we do not reject the hypothesis of affiliation in cost
signals.

This information is potentially useful to a policy maker. The apparent high
degree of estimated affiliation also explains why low levels of observed
competition are often sufficient to maintain relatively low profit margins:
strong affiliation is akin to fierce competition. Under strong
affiliation, a
potential winner knows that his nearest competitor probably has a valuation
(cost) close to his, and this disciplines his bidding behavior: he
becomes
more aggressive than under independence.

Our research has other policy implications, too. As mentioned above, it is
well known that, under affiliation, the English auction format, on average,
generates more revenue for the seller than the first-price, sealed-bid format.
In procurement, under affiliation, an English or a Vickrey auction would
get the job done more cheaply than the low-price, sealed-bid format. Were the
English or Vickrey formats being used and affiliation not rejected,
then the
procurement agency would be justified in its choice of mechanism. What remains
a bit of a puzzle is why the low-price, sealed-bid format is used in the
presence of such strong affiliation. Perhaps, other features, such as the
ability of the low-price, sealed-bid auction format to thwart collusion are
important, too. Alternatively, perhaps other moments of the bid distribution,
such as the variance, are important to the procurement agency.

On the other hand, had affiliation been rejected, then the procedures described
in Section \ref{s4} could be used to determine which auction format would get
the job
done most cheaply, on average. Again, it is possible that the English or
Vickrey formats would still be preferred. In any case, the methods described
in Section \ref{s4} permit a better understanding of the bidding differences, which
can aid in choosing the best auction format.

\section*{Acknowledgments}

Previous versions of this paper circulated under\break \citet{decapaar2007}
and \citet{decapaar2008}. For their comments on that research, we
thank Kurt
Anstreicher, Samuel Burer, Victor Chernozhukov, Srihari Govindan, Emmanuel
Guerre, Han Hong, Joel L.~Horowitz, Ali Horta\c csu, Kenneth L.~Judd, Roger
Koenker, David Prentice, Joseph P.~Romano, Che-Lin Su, Elie Tamer and
Michael Wolf. Timothy P.~Hubbard deserves special recognition and
thanks for
his advice and help with AMPL as well as his comments and insights on the
earlier research. We are also grateful to Stephen E.~Fienberg, Vijay Krishna,
Charles F. Manski, E. Glen Weyl and Frank A.~Wolak as well as two anonymous
referees who provided helpful comments and useful suggestions on the
penultimate draft of this paper.

\begin{supplement}
\stitle{Monte Carlo Study}
\slink[doi]{10.1214/00-AOAS344SUPP}
\slink[url]{http://lib.stat.cmu.edu/aoas/344/supplement.pdf}
\sdatatype{.pdf}
\sdescription{In this supplement, we discribe a small-scale Monte Carlo study used to investigate the numerical as well as small-sample properties of our testing strategy.}
\end{supplement}

\printaddresses

\end{document}